\documentclass[sigconf]{acmart}

\usepackage{subcaption,siunitx,booktabs}
\usepackage{multirow}
\usepackage{graphics}
\usepackage{graphicx}
\usepackage{ifthen}
\usepackage{xspace}
\usepackage{tablefootnote}
\usepackage{url}
\usepackage{balance}

\usepackage{filecontents}

\newboolean{showcomments}
\setboolean{showcomments}{true} 
\ifthenelse{\boolean{showcomments}}{
  \newcommand{\nbc}[3]{
    {\colorbox{#3}{\bfseries\sffamily\scriptsize\textcolor{white}{#1}}}%
    {\textcolor{#3}{\sf\small$\blacktriangleright$\textit{#2} $\blacktriangleleft$}}}
  \newcommand{\todo}[1]{\nbc{TODO}{#1}{red}\xspace}
  \newcommand{\fe}[1]{\nbc{Fe}{#1}{blue}\xspace}
  \newcommand{\becca}[1]{\nbc{Becca}{#1}{purple}\xspace}
}{
  \newcommand{\nbc}[3]{}
  \newcommand{\todo}[1]{}
  \newcommand{\fe}[1]{}
  \newcommand{\becca}[1]{}
}

 \newcommand{\sk}{{\sc Scikit-Learn}\xspace}
 \newcommand{\caret}{{\sc Caret}\xspace}
 \newcommand{\weka}{{\sc Weka}\xspace}
  \newcommand{\oob}{\textit{out-of-the-box-ml}\xspace}
   \newcommand{\tuned}{\textit{tuned-ml}\xspace}

\begin{document}

\title{Do Not Take It for Granted: Comparing Open-Source Libraries for Software Development Effort Estimation}

\author{Rebecca Moussa}
\affiliation{%
  \institution{University College London}
  \city{London} 
  \country{United Kingdom} 
}
\email{rebecca.moussa.18@ucl.ac.uk}

\author{Federica Sarro}
\affiliation{%
  \institution{University College London}
  \city{London} 
  \country{United Kingdom} 
}
\email{f.sarro@ucl.ac.uk}

\begin{abstract}
In the past two decades, several Machine Learning ({\sc ml}) libraries have become freely available.
Many studies have used such libraries to carry out empirical investigations on predictive Software Engineering (SE) tasks such as software development effort estimation, defect prediction and bug fixing time estimation. However, the differences stemming from using one library over another have been overlooked, implicitly assuming that using any of these libraries to run a certain machine learner would provide the user with the same or at least very similar results. 

This paper aims at raising awareness of the differences incurred when using different {\sc ml} libraries for software development effort estimation (SEE), which is one of most widely studied and well-known SE prediction tasks.

To this end, we investigate and compare four deterministic machine learners, widely used in SEE, as provided by three of the most popular {\sc ml} open-source libraries written in different languages (namely, \sk, \caret and \weka).

We carry out a thorough empirical study comparing the performance of these machine learners on the five largest available SEE datasets in the two most common SEE scenarios sought from the literature (i.e., \oob and \tuned) 
as well as an in-depth analysis of the documentation and code of their APIs.

The results of our empirical study reveal that the predictions provided by the three libraries differ in 95\% of the cases on average across a total of 105 cases studied. These differences are significantly large in most cases and yield misestimations of up to $\approx$ 3,000 hours ($\approx$ 18 months) per project. Moreover, our API analysis reveals that these libraries provide the user with different levels of control on the parameters one can manipulate, and a lack of clarity and consistency, overall, which might mislead users, especially less expert ones.

Our findings highlight that the {\sc ml} library is an important design choice for SEE studies, which can lead to a difference in performance. However, such a difference is under-documented. This suggests that end-users need to take such differences into account when choosing an {\sc ml} library, and developers need to improve the documentation of their {\sc ml} libraries to enable the end user to make a more informed choice.
We conclude our study by highlighting open-challenges with potential suggestions for the developers of these libraries as well as for the researchers and practitioners using them.

\end{abstract}



\maketitle

\section{Introduction}
As Machine Learning ({\sc ml}) becomes more pervasive and with the advent of several open-source ready-to-use libraries, {\sc ml} rapidly continues to be more accessible to a greater number of users of different levels of expertise. {\sc ml} techniques which were once hardly accessible to software engineers have now become widely available, thus shifting the task of realising fully hand-coded techniques to writing a few lines of API calls.

However, this democratization of {\sc ml} technology may come with some drawbacks. It has been observed that non-specialist users often revert to building out-of-the-box prediction models using these libraries, without properly experimenting with or investigating the extent to which this can influence their study/results \cite{kaltenecker2020interplay}. 
Similarly, previous studies have shown that it is not uncommon for software engineers to use such techniques without any modifications or enhancements \cite{tantithamthavorn2018impact}.

This paper aims at raising awareness of the use of different {\sc ml} libraries for Software Development Effort Estimation (SEE), a widely-known prediction problem in Software Engineering.

To this end we first reviewed 117 papers on SEE in order to understand the way in which the {\sc ml} libraries have been used in previous work \footnote{We examined 117 papers on {\sc ml} for SEE, which were included in two previous literature reviews published in 2012 \cite{wen2012systematic} and 2019 \cite{ali2019systematic}.} and found that 64\% of the studies (75 out of 117) run techniques without considering or exploring different values for their hyper-parameters (i.e., they use them out-of-the-box). Out of the 36\% that apply hyper-parameter tuning, 67\% (28 out of 42) of studies perform a manual trial and error, whereas the remaining studies (33\%) use a variety of techniques such as grid-search and search-based techniques (e.g., Genetic Algorithms and Tabu Search).

Therefore, we aim, herein, at addressing both of these scenarios: the first which depicts the use of machine learners out-of-the-box, where we investigate the extent to which building predictive models using the library API as-is (i.e., with default settings) would yield different results when using different libraries (i.e., \oob scenario), and the second which analyses the use of machine learners with hyper-parameter tuning (i.e., \tuned scenario) when using different libraries.

We investigate the above scenarios by carrying out a thorough empirical study with four deterministic {\sc ml} regression techniques (i.e., CART, KNN, LR and SVR) on the five largest publicly available datasets previously used in SEE studies~\cite{Sarro2018TOSEM, LFMTSE2020}, following best practice in empirical software prediction system assessments \cite{ShepperdM12}.

Moreover, we manually analyse the API documentation and the source code of the three libraries under study in order to unveil further characteristics.

The results of our \textit{empirical study} revealed that there is a large disagreement on the effort predictions achieved by a given machine learner using these three libraries (i.e., 95\% of the cases on average across all 105 cases studied). This indicates that the choice of the {\sc ml} library can contribute to the conclusion instability observed in previous SEE studies. 
Moreover, our analysis of 117 previous SEE work revealed that more than half do not state the {\sc ml} library used (51\%), and among those that do it, the majority only provides partial information (72\%), thus, making it difficult to reproduce and replicate previous work.

Furthermore, our \textit{analysis of the API documentation and code} reveals a lack of consistency among different libraries, which might not only lead to variance in the accuracy observed in our empirical study, but it might also induce users to misuse these APIs.

Overall our findings highlight that the choice of the {\sc ml} library, for a given SEE study, is just as important as that of predictors or evaluation measures, and we hope that the evidence provided herein will encourage future work to not only report the {\sc ml} library used for ease of reproducibility and replicability, but also to consider the choice of the library as an important design factor that can affect the results. These results also call for actions from the {\sc ml} developers to improve the clarity and consistency of the documentation of these libraries in order to enable the user to make an informed choice, and we provide some suggestions and future work towards this end.

To summarise, the contributions of our paper are:
\begin{itemize}
    \item [-] Raising awareness on the importance of the choice of {\sc ml} libraries for SSE.
     \item[-] Carrying out a thorough empirical study comparing four widely used SEE deterministic machine learners as provided by three very popular open-source {\sc ml} libraries on the five largest SEE datasets.
    \item[-] Providing an in-depth analysis of the APIs of these libraries aiming at investigating possible reasons for/sources of the large differences in accuracy observed in the empirical study.
    \item[-] Identifying some initial suggestions for both developers and users of {\sc ml} libraries on how to improve their documentation and usage and highlighting open-challenges to be addressed in future work.
\end{itemize}

\section{Related Work}
\label{sec:relatedwork}
The increasing growth and success of the use of machine learning has attracted software engineering researchers of various skill levels to use popular and free-of-charge {\sc ml} libraries to carry out Software Engineering (SE) tasks, including critical prediction tasks such as software development effort estimation \cite{BriandEnc:2002,wen2012systematic,ali2019systematic}, defect prediction \cite{hall2011systematic}, bug fixing time estimation \cite{BhattacharyaMSR11}. 

The number of such {\sc ml} libraries has been rapidly growing and their development is fast paced, leaving the engineer with a wide range of tools to choose from. General reviews of popular data analytics libraries can be found elsewhere \cite{MLDLToolsSurvey,bonthu2017review,dwivedi2016comprehensive}. 

If the results reported by these tools are largely consistent, then one can be confident in using any of these tools. However, previous work has shown that variances in results can arise from the use of different frameworks for deep learning \cite{doleck2020predictive,pham2020problems}, sentiment analysis \cite{Jongeling2017, Novielli18, Novielli21} and classification models in other application domains \cite{mitrpanont2017study, nandipati2020hepatitis,piccolo2020shinylearner,piccolo2021benchmarking}. 

Moreover, recent studies have highlighted that developers fail to grasp how to make {\sc ml}, and more generally {\sc ai}, libraries work properly, and very often seek further documentation or support from their peers on forums such as Stack Overflow \cite{islam2019developers,hashemi2020documentation}. Therefore, researchers have focused their attention on understanding potential problems faced by engineers when using these libraries \cite{begel2014analyze,amershi2019software,kechagia2022icse} and to ultimately improve, both the software documentation \cite{CummaudoTSE2020, hashemi2020documentation} and testing practices \cite{zhang2020machine,CynthiaOracleTestingML} for {\sc ml}/{\sc ai} software. 

To the best of our knowledge, there is no previous work that investigates differences arising from the use of {\sc ml} libraries for predictive modelling in software engineering, precisely software effort estimation. 
While previous studies in effort estimation have shown that differences in training data, learning techniques, hyper-parameter tuning and evaluation procedures can lead to variance in the prediction result causing conclusion instability \cite{Myrtveit05,Menzies2012,Song13,Keung2013,Sarro18,SarroTSE20}, no study has investigated the impact that the use of different {\sc ml} tools to build the prediction model can have on the variance in the estimates. Differently from previous work, which has mainly investigated the variance of non-stochastic approaches \cite{pham2020problems,liem2020run}, we intentionally design an empirical study in a way that reduces, as much as possible, all those factors that can introduce stochasticity in the machine learner output. This allows us to analyse the variances in performance of a given machine learner, which are primarily due to the use of the different {\sc ml} libraries investigated in our work.

\section{Empirical Study}  
\label{sec:emp-design}
In this section, we describe the design and results of the empirical study we have carried out to assess the extent to which using machine learners provided by different {\sc ml} libraries might yield different results.

\subsection{Design}

\subsubsection{Research Questions}
\label{sec:RQsdesign}
The first step towards identifying whether there are any discrepancies in prediction accuracy performance when using different {\sc ml} libraries for SEE lies in recognizing the frequency of its occurrence, which motivates our first research question: 

\noindent \textbf{RQ1. Prediction Accuracy:} \textit{How often a given machine learner provides different SEE results when built by using different {\sc ml} libraries?}

If we found that, in fact, the results differ in many cases, then it is important to verify the way in which the SEE prediction performance of a machine learner is affected depending by the {\sc ml} library used. Thus our second research question assesses: 

\noindent \textbf{RQ2. Change in Prediction Performance:} \textit{How does the SEE performance of a given machine learner change when built using different {\sc ml} libraries?}

In other words, while RQ1 aims at revealing whether the number of cases showing inconsistency in results due to the {\sc ml} library used is high, RQ2 further analyses these results by looking into the magnitude of these differences to understand if they are relevant. 

Lastly, we aim at shedding light on questions like "\textit{Would my conclusion have been different had I used \sk instead of \caret?}", \textit{"Would the results of my proposed approach have been better or worse, compared to others, had another {\sc ml} library been used?"}. As previous study usually ranks machine learners based on their estimation performance for comparison purposes, our third and last research question asks:

\textbf{RQ3. Change in Ranking:} \textit{How does the ranking of SEE machine learners change when considering different {\sc ml} libraries?}

If we find that different rankings are provided by different {\sc ml} libraries, this would point out possible conclusion instability threats in previous (and future) studies given that, for example, a technique that would rank first in a study using a given library, might not have the same rank (i.e., first) when another library is used instead.

\subsubsection{{\sc ml} Libraries}
\label{sec:API}
In our empirical study we compare three of the most popular open-source {\sc ml} libraries written in different languages: \sk \cite{pedregosa2011scikit} in {\sc Python}, \caret \cite{kuhn2015short} in {\sc R}, and \weka \cite{weka} in Java.

\sk is a Python library, distributed under BSD license, which includes a wide range of state-of-the-art supervised and unsupervised machine learning techniques. 

The aim of the \sk is to provide efficient and well-established machine learning libraries within a programming environment that is accessible to non-machine learning experts and reusable in various scientific areas \cite{buitinck2013api}.

\sk is designed to adhere to a number of engineering principles, including the use of sensible defaults stating ``Whenever an operation requires a user-defined parameter, an appropriate default value is defined by the library. The default value should cause the operation to be performed in a sensible way.'' 

\caret, acronym for Classification and Regression Training, is an {\tt R} package, distributed under the GNU General Public Licence, containing functions to streamline model training process for complex regression and classification problems \cite{kuhn2015short}. It was made publicly available on CRAN in 2007 and it relies on several other {\tt R} packages that are not installed at the package start-up, but loaded as needed. 

The aim of \caret is to provide the user with an easy interface for the execution of several classifiers, allowing automatic parameter tuning and reducing the requirements on the researcher’s knowledge about the tunable parameter values, among other issues \cite{kuhn2008building}.

\weka, acronym for Waikato Environment for Knowledge Analysis, is licensed under the GNU General Public Licence and was first released in 1997 \cite{weka}.
\weka provides implementations of machine learning techniques that can be easily used through a simplified and interactive graphical interface by users who cannot or do not need to write code, or through an API that allows user to access the libraries from their own Java programs. The online documentation is automatically generated from the source code and concisely reflects its structure \cite{h2016data}. 
\weka provides the main methods supervised and unsupervised machine learning techniques, as well as methods for data pre-processing and visualization. 

In our empirical study, we use the latest stable version for each library at the time the study was done: \sk  0.23.1, \caret  6.0.85, \weka 3.8 with {\tt Python} 3.7.5, {\tt R 3.6.3}, and {\tt Java 11}, respectively.

\subsubsection{Machine Learners and Settings}
\label{subsec:techniques}
In this section we briefly introduce the machine learners we investigated in order to compare the performance of the three {\sc ml} libraries \caret, \sk and \weka. We specifically focus on a set of deterministic {\sc ml} techniques widely used in SEE in order to eliminate any randomness arising from the stochastic nature of other machine learners, and thereby to analyse differences stemming only from the library usage. These techniques are Classification and Regression Tree (CART) \cite{cart84}, K-Nearest Neighbours (KNN) \cite{ripley2007pattern}, Linear Regression (LR) \cite{seber2012linear} and Support Vector Regression (SVR) \cite{burges1998tutorial}.

In Table \ref{tab:default-param}, we list these machine learners together with the name of the class implementing each of them within each {\sc ml} library studied herein.

\begin{table}[]
\caption{Machine learners investigated and corresponding class/method name in \sk, \caret and \weka.}
\resizebox{0.75\columnwidth}{!}{
\begin{tabular}{l|ll}
\hline
\multicolumn{1}{c|}{\textbf{Machine Learner}} & \multicolumn{1}{c}{\textbf{Library}} & \multicolumn{1}{c}{\textbf{Class/Method Name}}                                                                                                  \\ \hline
\multirow{3}{*}{\textbf{CART}}           

& Caret                             & rpart                                                                                                                                                                                            \\ \cline{2-3} 
& SkLearn                           & DecisionTreeRegressor                 \\ \cline{2-3} 
                                          
                                          & Weka                              & REPTree                                                                                                                                                                                                                                                                                                                                \\ \hline
\multirow{3}{*}{\textbf{KNN}}           
& Caret                             & knn                                                                                                                                                                                                                                                                                                                                                             \\ \cline{2-3} 

& SkLearn                           & KNeighborsRegressor                                                                                                                                                                                 \\ \cline{2-3} 
                                          
                                          & Weka                              & IBk                                                                        \\ \hline
\multirow{3}{*}{\textbf{LR}}            
 & Caret                             & lm                                                                                                                     \\ \cline{2-3} 

& SkLearn                           & LinearRegression                                                                          \\ \cline{2-3} 
                                         
                                          & Weka                              & SimpleLinear                                                                                                                                                                                                                                                       \\\hline
\multirow{3}{*}{\textbf{SVR}}          
& Caret                             & svmRadial                                                                  \\ \cline{2-3} 
& SkLearn                           & SVR                                                                                                               \\ \cline{2-3} 
                                          
                                          & Weka                              & SMOReg                                                          \\ \hline
\end{tabular}}
\label{tab:default-param}
\end{table}

To investigate the \oob scenario, we use all the techniques with the default settings provided by each of the {\sc ml} libraries, without applying any further modifications. 

To investigate the \tuned scenario, we apply Grid Search as a hyper-parameter tuning technique. We chose Grid Search over other tuning technique (e.g., Random Search) in order to eliminate any difference in performance resulting from external factors (i.e., any stochastic behaviour or anything other than the {\sc ml} libraries themselves). This allows us to accurately verify whether variances in performance result from the {\sc ml} libraries or from other factors like from using Random Search for hyper-parameter tuning, for example. 
We run each tool's corresponding grid search method 30 times (i.e., {\tt GridSearchCV} in \sk, {\tt tuneGrid} along with {\tt trControl} in \caret and {\tt GridSearch} in \weka) using the same inner cross-validation across all tools to minimize any possible stochastic behaviour deriving from the data splits and using the same set of values for the hyper-parameters across all three libraries. In order to maintain a fair comparison, we tune the parameters that are found in common across all three libraries. The settings used for both the \oob and \tuned scenarios can be found in our on-line appendix \cite{onlineappendix}.

\subsubsection{Datasets}
To empirically investigate our RQs we used the largest SEE publicly available datasets (namely, China, Desharnais, Kitchenam, Maxwell, Miyazaki) containing a diverse sample of industrial software projects developed by a single company or several software companies \cite{seacraftrepo}.
These datasets exhibit a high degree of diversity both in terms of number of observations (from 48 to 499), number and type of features (from 3 to 17), technical characteristics (e.g., software projects developed in different programming languages and for different application domains), number of companies involved and their geographical locations. Furthermore, all these datasets have been widely used in several SEE studies (see e.g., \cite{Sarro:2016,Sarro2012MOGA,Ferrucci:2014sbpm,KocaguneliEnsemble,sigweni2016,Sarro2018TOSEM,Shepperd:2000,LFMTSE2020}). 

A comprehensive description of these datasets, together with the actual data is available on-line \cite{onlineappendix}, while in Table \ref{tab:descriptiveStatisticsDatasets} we report for each of the datasets its type (Within-Company --WC-- or Cross-Company --CC-- \cite{WCCCEEsurveyEmilia}), number of projects, and the descriptive statistics of the dependent variable (i.e., Effort) and the independent variables used to build the prediction models.

\begin{table}[]
\centering
\caption{Datasets used in our study.}

\resizebox{0.85\columnwidth}{!}{
\begin{tabular}{lllrrrr}
  \hline
 Dataset & Type & Variable & Min & Max & Mean & Std. Dev. \\ 
  \hline

  \hline
	China & CC& Input & 0.00 & 9404.00 & 167.10 & 486.34\\
	(499 projects)      & & Output & 0.00 & 2455.00 & 113.60 & 221.27\\
	      & & Enquiry & 0.00 & 952.00 & 61.60 & 105.42 \\
	      & & File & 0.00 & 2955.00 & 91.23 & 210.27\\
	      & & Interface & 0.00 & 1572.00 & 24.23 & 85.04\\
	      & & Effort & 26.00 & 54620 .00& 3921.00 & 6481.00\\
	\hline
	Desharnais &WC & TeamExp & 0.00 & 4.00 & 2.30 & 1.33\\
	(77 projects)		   & & ManagerExp & 0.00 & 4.00 & 2.65 & 1.52\\
			   & & Entities & 7.00 & 386 & 121.54 & 86.11\\
			   & & Transactions & 9.00 & 661.00 & 162.94 & 146.09\\
			   & & AdjustedFPs & 73.00 & 1127.00 & 284.48 & 182.26\\
			   & & Effort & 546.00 & 23490.00 & 4903.95 & 4188.19\\
	\hline

  Kitchenham & CC& AFP & 15.36 & 18140 & 527.70 & 1521.99 \\
	(145 projects)		 & & Effort & 219.00 & 113900.00& 3113.00 & 9598.00 \\   
  \hline
	Maxwell & CC & SizeFP & 48.00 & 3643.00 & 673.31 & 784.04\\
	(62 projects)	& & Nlan & 1.00 & 4.00 & 2.55 & 1.02\\
			& & T01 & 1.00 & 5.00& 3.05 & 1.00\\
			& & T02 & 1.00 & 5.00 & 3.05 & 0.71\\
			& & T03 & 2.00 & 5.00 & 3.02 & 0.89\\
			& & T04 & 2.00 & 5.00 & 3.19 & 0.70\\
			& & T05 & 1.00 & 5.00 & 3.05 & 0.71\\
			& & T06 & 1.00 & 4.00 & 2.90 & 0.69\\
			& & T07 & 1.00 & 5.00 & 3.24 & 0.90\\
			& & T08 & 2.00 & 5.00 & 3.81 & 0.96\\
			& & T09 & 2.00 & 5.00 & 4.06 & 0.74\\
			& & T10 & 2.00 & 5.00 & 3.61 & 0.89\\
			& & T11 & 2.00 & 5.00 & 3.42 & 0.98\\
			& & T12 & 2.00 & 5.00 & 3.82 & 0.69\\
			& & T13 & 1.00 & 5.00 & 3.06 & 0.96\\
			& & T14 & 1.00 & 5.00 & 3.26 & 1.01\\
			& & T15 & 1.00 & 5.00 & 3.34 & 0.75\\
			& & Effort & 583.00 & 63694.00 & 8223.20 & 10500.00\\
\hline
			Miyazaki & CC& SCRN & 0.00 & 281.00 & 33.69 & 47.24\\
		   (48 projects)		 & & FORM & 0.00 & 91.00 & 22.38 & 20.55\\
			 & & FILE & 2.00 & 370.00 & 34.81 & 53.56\\
			 & & Effort & 896.00 & 253760.00 & 13996.00 & 36601.56\\
\hline
\hline
\end{tabular}}
\label{tab:descriptiveStatisticsDatasets}
\end{table}

\subsubsection{Validation}
In order to eliminate any possible variance in performance caused by non-deterministic influencing factors such as the validation approach employed or the sampled data used as denoted by Rahman et al. \cite{rahman2013sample}, we perform a Leave-One-Out cross-validation (LOO), where each instance (i.e., a software project in our case) of a dataset of $n$ observations is considered to be a fold. At each iteration, the learning process includes training a prediction model on $n-1$ instances and testing it on the one instance left out. LOO is a deterministic approach that, unlike other cross validation techniques, does not rely on any random selection to create the training and testing sets. Therefore, this validation process is fully reproducible and eliminates conclusion instability caused by random sampling \cite{kocaguneli2013software, LFMTSE2020}. However, it has been observed that LOO may give more optimistic results than those that might realistically be achieved in practice \cite{sigweni2016realistic} and if chronological information about the projects is available it would be preferable to adopt a time-based validation approach \cite{sigweni2016realistic,LFMTSE2020}. In our study we use LOO because start and completion dates are not available for all projects.

\subsubsection{Evaluation Measures}
\label{sec:evaluationCriteria}
Several measures have been proposed to evaluate the accuracy of effort estimation prediction models. They are generally based on the \textit{absolute error}, i.e., $|\mbox{\emph{ActualEffort}} - \mbox{\emph{EstimatedEffort}}|$.

In our study we use the \emph{Mean Absolute Error} (MAE) as it is unbiased towards both over- and under-estimation \cite{ShepperdM12,Langdon2016}. Given a set of $N$ projects and the measured ($ActualEffort_i$) and estimated ($EstimatedEffort_i$) effort for each of the the project $i$ in this set, MAE is defined as follows: 
$MAE = \frac{1}{N} \sum_{i=1}^{N}  |ActualEffort_i - EstimatedEffort_i|$.

We also use statistical significance tests to assess differences in the performance of the {\sc ml} libraries (RQ2) and the way they do rank machine learners (RQ3).

In order to evaluate whether the differences in MAE values resulting from the use of various {\sc ml} libraries are statistically significant (RQ2), we perform the Wilcoxon Signed-Rank Test \cite{woolson2007wilcoxon} at a confidence limit, $\alpha$, of 0.05, with Bonferroni correction ($\alpha/K$, where $K$ is the number of hypotheses) when multiple hypothesis are tested. 
Unlike parametric tests, this test does not make any assumptions about underlying data distributions. The null hypothesis that is tested in our work follows: \textit{"There is no significant difference in the MAE values obtained by the approaches when built using a different {\sc ml} library"}.
We also compute the Vargha and Delaney's $\hat{A}_{12}$ non-parametric effect size measure to assess whether any statistically significant difference is worthy of practical interest~\cite{ArcuriB14}. $\hat{A}_{12}$ is computed based on the following formula $\hat{A}_{12} = (R_1/m - (m + 1)/2)/n$, where $R_1$ is the rank sum of the first data group we are comparing, and $m$ and $n$ are the number of observations in the first and second data group, respectively. When the two compared groups are equivalent $\hat{A}_{12} = 0.5$. An $\hat{A}_{12}$ higher than $0.5$ denotes that the first data group is more likely to produce better results. The effect size is considered small when $0.5 < \hat{A}_{12} \leq 0.66$, medium when $0.66 < \hat{A}_{12} \leq 0.75$ and large when $\hat{A}_{12}>0.75$, although these thresholds are not definitive~\cite{Sarro:2016}. 

In order to verify statistical significance, in the ranking provided by the  \sk, \caret and \weka APIs (RQ3), we use the Friedman Test \cite{friedman1937use}. This is a non-parametric test which works with the ranks of the data groups rather than their actual performance values, making it less susceptible to the distribution of the performance of these parametric values. When a significant difference is found, the Nemenyi test \cite{nemenyi1963distribution} is often recommended as a post-hoc test to identify the data groups with a statistically significant difference \cite{demvsar2006statistical}. The performance of two data groups is thought to be statistically significantly different if the corresponding average ranks differ by at least the Critical Distance (CD) \cite{demvsar2006statistical}. The results of this test are presented in a diagram which is used to compare the performance of multiple techniques by ranking them based on each {\sc ml} library. It consists of an axis, on which the average ranks of the methods are plotted and of the CD bar. The CD bars of the groups of classifiers whose values are significantly different do not overlap. The visualisation used herein is a more recent version of the one presented in Demsar’s work \cite{demvsar2006statistical}, which aims at providing an easier interpretation. It can be obtained by using the {\tt Nemenyi} function in {\sc R}.

\subsection{Results}
\label{sec:results}
In this Section we report the results obtained in our empirical study for each of the RQs explained in Section \ref{sec:RQsdesign}.

\subsubsection{RQ1: Prediction Results} 
\label{sec:resultsRQ1}

Our first research question asks how often a certain prediction model built using different {\sc ml} libraries for SEE provides same results. In order to answer this question, we look at the cases where {\sc ml} libraries result in different predictions. As described in Section \ref{sec:emp-design}, we explore two common scenarios observed in the SEE literature, the use of out-of-the-box prediction models (i.e., \oob) and their use when hyper-parameter tuning is performed (i.e., \tuned). 

Tables \ref{tab:out-of-box-error} and \ref{tab:tuning-error} show the results for each scenario, respectively. As one can observe from Table \ref{tab:out-of-box-error} (i.e., \oob scenario), there are only five out of the 60 cases under study (8\%) where two out of the three {\sc ml} libraries provide the same predictions. Specifically, LR achieves the same outcome when built using \sk and \caret, whereas \weka obtains different results. Differences in prediction are observed for all other techniques when built using any of the three {\sc ml} libraries. 

As for the second scenario, we explore prediction models which are tune-able (i.e., they consist of parameters for which hyper-parameter tuning can be applied): CART, KNN and SVM.\footnote{We exclude LR as there are no tune-able parameters common to all libraries} Table \ref{tab:tuning-error} shows that there is a difference in prediction in all of the cases studied (i.e., there is no case where all three libraries agree on the same prediction).    

\begin{table*}[tb]
\caption{RQ1-RQ2: Differences in MAE values obtained when building prediction models using \sk (Sk), \caret (C) and \weka (W) in the (a) \oob scenario and the (b) \tuned scenario. The $\uparrow$ indicates that the $1^{st}$ tool produces worse predictions than the $2^{nd}$, whereas the $\downarrow$ indicates that $1^{st}$ tool produces better predictions than the $2^{nd}$.}
\begin{subtable}{1\textwidth}

\centering
  \resizebox{0.9\linewidth}{!}{
\begin{tabular}{llll|lll|lll|lll}
\hline
           & \multicolumn{3}{c}{CART}                                                                                                                   & \multicolumn{3}{c}{KNN}                                                                                                                                           & \multicolumn{3}{c}{LR}                                                                    & \multicolumn{3}{c}{SVM}                                                                    \\
           & Sk vs. C                                             & Sk vs. W                     & C vs. W                                              & Sk vs. C                                            & Sk vs. W                                             & C vs. W                                              & Sk vs. C                  & Sk vs. W                      & C vs. W                       & Sk vs. C                     & Sk vs. W                     & C vs. W                      \\ \hline
China      &  $\downarrow$643                          &  $\downarrow$477  &  $\uparrow$166                         &  $\downarrow$99                          &   $\uparrow$734  &    $\uparrow$833  & 0                 &  $\uparrow$420  &  $\uparrow$420  &  $\downarrow$359  &  $\downarrow$637  &  $\downarrow$278  \\
Desharnais &  $\downarrow$396                          &  $\uparrow$172 &  $\uparrow$568                         &  $\uparrow$111                        &  $\uparrow$972  &  $\uparrow$861                         & 0                 &  $\downarrow$226   &  $\downarrow$226   &  $\downarrow$440  &  $\downarrow$577  & $\downarrow$137  \\
Kitchenham &  $\downarrow$46                           &  $\downarrow$227  &  $\downarrow$181                          &  $\downarrow$36                          &  $\uparrow$125  &  $\uparrow$161  & 0                 &  $\uparrow$ 177  &  $\uparrow$ 178  &  $\downarrow$134  &  $\downarrow$683  &  $\downarrow$550  \\
Maxwell    &  $\downarrow$3079                         &  $\downarrow$1095    &  $\uparrow$1983 &  $\uparrow$193 &  $\uparrow$2435 &  $\uparrow$2242 & 0                 &  $\downarrow$450   &  $\downarrow$450   &  $\downarrow$1591 &  $\downarrow$1319 &  $\uparrow$272 \\
Miyazaki   &  $\uparrow$1652 &  $\downarrow$578  &  $\downarrow$2231                         &  $\uparrow$75  &  $\uparrow$518  &  $\uparrow$443  & 0                 &  $\uparrow$1229 &  $\uparrow$1229 &  $\downarrow$714  &  $\downarrow$3188 &  $\downarrow$2474 \\ \hline
\end{tabular}
}
\caption{\oob scenario.}
\label{tab:out-of-box-error}
\end{subtable}

\begin{subtable}{1\textwidth}
\centering
  \resizebox{0.7\linewidth}{!}{
\begin{tabular}{llll|lll|lll}
\hline
           & \multicolumn{3}{c}{CART}           & \multicolumn{3}{c}{KNN}                                &  \multicolumn{3}{c}{SVM}                 \\
           & Sk vs. C                                              & Sk vs. W                       & C vs. W                                      & Sk vs. C                  & Sk vs. W                       & C vs. W                        & Sk vs. C                      & Sk vs. W                      & C vs. W                       \\ \hline
China      & $\downarrow$809                           & $\downarrow$560    & $\uparrow$249                          & $\uparrow$10                          & $\downarrow$6    & $\downarrow$16        & $\downarrow$421   & $\downarrow$598   & $\downarrow$177   \\
Desharnais & $\downarrow$1,113                         & $\downarrow$457    & $\uparrow$656                          & $\downarrow$29                           & $\uparrow$160 & $\uparrow$189                               & $\downarrow$336   & $\downarrow$419   & $\downarrow$83    \\
Kitchenham & $\uparrow$185                          & $\uparrow$247   & $\uparrow$62                           & $\uparrow$14                          & $\downarrow$25   & $\downarrow$39       & $\downarrow$250   & $\downarrow$237   & $\uparrow$14   \\
Maxwell    & $\downarrow$2,515                         & $\downarrow$652    & $\uparrow$1,864 & $\downarrow$421   & $\uparrow$668 & $\uparrow$1,089   & $\downarrow$1,044 & $\downarrow$1,738 & $\downarrow$694   \\
Miyazaki   & $\uparrow$3,053 & $\uparrow$1,872 & $\downarrow$1,181                         & $\downarrow$1,104 & $\downarrow$401  & $\uparrow$703  & $\downarrow$966   & $\downarrow$2,232 & $\downarrow$1,266 \\ \hline
\end{tabular}}
  \caption{\tuned scenario.}\label{tab:tuning-error}
\end{subtable}
\end{table*}

\subsubsection{RQ2: Change in Prediction Performance}
\label{sec:resultsRQ2}
 
Our second research question investigates how the performance of a given SEE technique changes when the prediction model is built using different {\sc ml} libraries. 

Tables \ref{tab:out-of-box-error} and \ref{tab:tuning-error} show the difference in the MAE results for each technique obtained by comparing each pair of the {\sc ml} libraries under study for the two scenarios analysed. We denote any positive difference by the up arrow, whereas any negative difference is represented by the down arrow. For each pair of tools, a positive difference signifies that the {\sc ml} library listed as first achieves a higher MAE value (i.e., a less accurate prediction) than the other library. On the other hand, a negative difference denotes that the first library  provides an MAE value lower than the one provided by the second library (i.e., it provides a better prediction).
For example, when considering CART, specifically comparing it when it is built using \sk and \caret (i.e., denoted as \textit{Sk vs. C}) the difference in MAE is $\uparrow$ 809 meaning that when CART is built using \sk, it achieved a higher MAE (i.e., worse prediction performance) than that obtained when CART is built using \caret with a difference of 809 man-hours. 

In order to study the magnitude of the error between the {\sc ml} libraries, we analyse our results using three error ranges: a difference of 100, 500 and 1,000 hours. By doing so we relax our constraint on what we consider the difference to be impactful by allowing/accepting a larger error between the results obtained by the libraries.

Table \ref{tab:both-tables} present the number of times the difference in results falls within these three ranges. 

When considering the \oob prediction scenario, we can observe that while \sk and \caret provide the same MAE for LR only, they, along with \weka achieve different results in all other cases (i.e., 51 out of the 60) with a difference of at least 100 hours. As shown in Table \ref{tab:both-tables}, the magnitude of the discordance in performance is very large for every pair of {\sc ml} library under comparison with 22\% of cases reaching a difference of at least a 1,000 hours. 

When analysing the MAEs using a threshold $\geq$ 100 hours, results show that \sk and \caret are the least discordant pair with more than half the cases (55\%) having a difference of at least 100 hours. When comparing \sk with \weka and \caret with \weka, a difference $\geq$ 100 is seen in all the cases considered. Moreover, we also found cases with a difference of at least 500 hours. 

Specifically, between  \sk and \caret there is a difference $\geq$ 500 hours in 25\% of the cases (5 out of 20), whereas between \caret and \weka it exists in 45\% of the cases  and between \sk and \weka, it is in 60\% of the cases (12 out of 20). 

One would not expect to obtain a difference of such a high magnitude resulting from the use of one {\sc ml} library over another, however these cases exist even with a difference $\geq$ 1,000 hours. As we can observe from Table \ref{tab:both-tables}, when comparing the results obtained by \sk with those by \caret, there is a difference of at least 1,000 hours in 15\% of the cases under study. As for the differences between the other two pair of libraries (i.e., \sk vs. \weka and \caret vs. \weka), this difference is seen in 25\% of the cases.

When considering the \tuned scenario, we observe that differences occur more frequently  than in the \oob scenario. We found a difference of at least 100 hours in 80\% of the cases (12 out of 15) for \sk vs. \caret, in 87\% of the cases (13 out of 15) for \sk vs. \weka and in 67\% of the cases (10 out of 15) for \caret vs. \weka (Table \ref{tab:both-tables}). 
Several differences still persist when we consider a difference of at least 500 hours: specifically in 47\% of the cases (7 out of 15) when comparing each pair of libraries  (i.e., \sk vs. \caret, \sk vs. \weka and \caret vs. \weka). By considering any difference less than a 1,000 hours to be negligible, we observe that differences $\geq$ 1,000 still exist in 33\% of the cases for \sk vs. \caret, in 20\% of the cases for \sk vs. \weka, and in 27\% for \caret vs. \weka.

These results reveal that differences as large as or even larger than 1,000 hours still exist for almost all techniques. Such magnitude in difference would not only have an impact on a project's feasibility and completeness but can also change the conclusion of any study proposing or comparing a new or existing techniques as we further investigate in RQ3.

The statistical significance test results show that when comparing \sk and \caret, the differences in MAE values still prove to be statistically significantly different (p-value < 0.01 and A12 = 0.55) despite the low statistical power due to the small sample size being tested. The same observation could not be made when comparing \sk and \weka (p-value = 0.164), and \caret and \weka (p-value = 0.310) even though the MAE results reported in Table \ref{tab:rawmae} show a large variance between the three {\sc ml} libraries (for three out of the four techniques investigated). This could be due to the small sample size which can result in a \textit{Type II error}, where the test fails to reject the null hypothesis, even though it should not be the case. 

In order to further understand the severity of the error, we also compute the relative deviation in effort estimation with respect to the mean actual effort of the projects in a given dataset (see Table \ref{tab:descriptiveStatisticsDatasets}). For example, if the average actual effort spent for realising projects in a company is 10,000 hours then it is less severe to have a deviation $\geq$ 1,000 hours than with an average actual effort of 50 hours. We observe that for the \oob scenario, the largest differences between {\sc ml} libraries range from 20\% to 37\% depending on the dataset. For example, we can observe a difference of 3,188 hours in the results achieved by \sk and \weka on the Miyazaki dataset, which corresponds to a 23\% deviation in effort estimation with respect to the actual effort. As for the \tuned scenario, we can observe that the deviation ranges between 8\% and 31\%, with four out of the five datasets having a deviation greater than 20\%.

\begin{table}[]
\caption{RQ2: Number of cases the results provided by a given {\sc ml} library differ from another of at least 100 hours, 500 hours and 1,000 hours for the \oob scenario and the \tuned scenario.} 
\resizebox{1\columnwidth}{!}{
\begin{tabular}{llllll|llll}
\hline
                                                                                               & \multicolumn{5}{c|}{\textit{out-of-the-box-ml}} & \multicolumn{4}{c}{\textit{tuned--ml}} \\ \hline
\multicolumn{1}{l|}{\textit{\begin{tabular}[c]{@{}l@{}}Prediction \\ Difference\end{tabular}}} & CART     & KNN    & LR    & SVM    & Overall    & CART     & KNN    & LR    & Overall    \\ \hline
\textit{}                                                                                      & \multicolumn{5}{c|}{Sklearn vs. Caret}          & \multicolumn{4}{c}{Sklearn vs. Caret}  \\ \hline
\multicolumn{1}{l|}{\textit{$\geq$ 100 h}}                                                     & 4        & 2      & 0     & 5      & 11         & 5        & 2      & 5     & 12         \\
\multicolumn{1}{l|}{\textit{$\geq$ 500 h}}                                                     & 3        & 0      & 0     & 2      & 5          & 4        & 1      & 2     & 7          \\
\multicolumn{1}{l|}{\textit{$\geq$ 1,000 h}}                                                   & 2        & 0      & 0     & 1      & 3          & 3        & 1      & 1     & 5          \\ \hline
                                                                                               & \multicolumn{5}{c|}{Sklearn vs. Weka}           & \multicolumn{4}{c}{Sklearn vs. Weka}   \\ \hline
\multicolumn{1}{l|}{\textit{$\geq$ 100 h}}                                                     & 5        & 5      & 5     & 5      & 20         & 5        & 3      & 5     & 13         \\
\multicolumn{1}{l|}{\textit{$\geq$ 500 h}}                                                     & 2        & 4      & 1     & 5      & 12         & 3        & 1      & 3     & 7          \\
\multicolumn{1}{l|}{\textit{$\geq$ 1,000 h}}                                                   & 1        & 1      & 1     & 2      & 5          & 1        & 0      & 2     & 3          \\ \hline
                                                                                               & \multicolumn{5}{c|}{Caret vs. Weka}             & \multicolumn{4}{c}{Caret vs. Weka}     \\ \hline
\multicolumn{1}{l|}{\textit{$\geq$ 100 h}}                                                     & 5        & 5      & 5     & 5      & 20         & 4        & 3      & 3     & 10         \\
\multicolumn{1}{l|}{\textit{$\geq$ 500 h}}                                                     & 3        & 3      & 1     & 2      & 9          & 3        & 2      & 2     & 7          \\
\multicolumn{1}{l|}{\textit{$\geq$ 1,000 h}}                                                   & 2        & 1      & 1     & 1      & 5          & 2        & 1      & 1     & 4          \\ \hline
\end{tabular}}
\label{tab:both-tables}
\end{table}

\subsubsection{RQ3: Change in Ranking}
\label{sec:resultsRQ3}

In order to further understand the magnitude of the difference and to investigate whether it can have an impact on a study's conclusion, we analyse the ranking of the techniques, based on the MAE achieved, according to each library separately and assess the differences.

\begin{table*}[tb]
\caption{RQ3: Rankings of prediction models based on the MAE results obtained by each of the {\sc ml} libraries for each of the five datasets for the (a) \oob scenario and the (b) \tuned scenario.} 
\begin{subtable}{1\textwidth}
\centering
\resizebox{0.88\linewidth}{!}{
\begin{tabular}{lll|lll|lll|lll|lll}
\hline
\multicolumn{3}{c|}{\textbf{China}}                                                                            & \multicolumn{3}{c|}{\textbf{Desharnais}}                                                                       & \multicolumn{3}{c|}{\textbf{Kitchenham}}                                                                       & \multicolumn{3}{c|}{\textbf{Maxwell}}                                                                          & \multicolumn{3}{c}{\textbf{Miyazaki}}                                                                         \\ 
\multicolumn{1}{l}{\textbf{SkLearn}} & \multicolumn{1}{l}{\textbf{Caret}} & \multicolumn{1}{l|}{\textbf{Weka}} & \multicolumn{1}{l}{\textbf{SkLearn}} & \multicolumn{1}{l}{\textbf{Caret}} & \multicolumn{1}{l|}{\textbf{Weka}} & \multicolumn{1}{l}{\textbf{SkLearn}} & \multicolumn{1}{l}{\textbf{Caret}} & \multicolumn{1}{l|}{\textbf{Weka}} & \multicolumn{1}{l}{\textbf{SkLearn}} & \multicolumn{1}{l}{\textbf{Caret}} & \multicolumn{1}{l|}{\textbf{Weka}} & \multicolumn{1}{l}{\textbf{SkLearn}} & \multicolumn{1}{l}{\textbf{Caret}} & \multicolumn{1}{l}{\textbf{Weka}} \\ \hline
LR      & LR& SVM                                & KNN & LR      & LR      & LR      & LR      & SVM                                & KNN       & KNN                                & LR      & KNN       & KNN                                & SVM                               \\
KNN       & KNN                                & LR      & LR      & KNN                                & SVM                                & KNN       & KNN                                & LR      & LR      & SVM                                & SVM                                & SVM       & SVM                                & LR                                \\
SVM       & SVM                                & CART                               & SVM       & SVM                                & CART                               & SVM       & SVM                                & KNN                                & SVM       & LR      & KNN                                & LR      & LR      & CART                              \\
CART      & CART                               & KNN                                & CART      & CART                               & KNN                                & CART      & CART                               & CART                               & CART      & CART                               & CART                               & CART      & CART                               & KNN                               \\ \hline
\end{tabular}}
\caption{\oob scenario.}\label{tab:out-of-the-box-ranking}
\end{subtable}

\bigskip

\begin{subtable}{1\textwidth}
\centering
\resizebox{0.88\linewidth}{!}{
\begin{tabular}{lll|lll|lll|lll|lll}
\hline
\multicolumn{3}{c|}{\textbf{China}}                                                                            & \multicolumn{3}{c|}{\textbf{Desharnais}}                                                                       & \multicolumn{3}{c|}{\textbf{Kitchenham}}                                                                       & \multicolumn{3}{c|}{\textbf{Maxwell}}                                                                          & \multicolumn{3}{c}{\textbf{Miyazaki}}                                                                         \\
\multicolumn{1}{l}{\textbf{SkLearn}} & \multicolumn{1}{l}{\textbf{Caret}} & \multicolumn{1}{l|}{\textbf{Weka}} & \multicolumn{1}{l}{\textbf{SkLearn}} & \multicolumn{1}{l}{\textbf{Caret}} & \multicolumn{1}{l|}{\textbf{Weka}} & \multicolumn{1}{l}{\textbf{SkLearn}} & \multicolumn{1}{l}{\textbf{Caret}} & \multicolumn{1}{l|}{\textbf{Weka}} & \multicolumn{1}{l}{\textbf{SkLearn}} & \multicolumn{1}{l}{\textbf{Caret}} & \multicolumn{1}{l|}{\textbf{Weka}} & \multicolumn{1}{l}{\textbf{SkLearn}} & \multicolumn{1}{l}{\textbf{Caret}} & \multicolumn{1}{l}{\textbf{Weka}} \\ \hline
KNN       & KNN    & SVM   & KNN        & KNN      & SVM     & KNN        & KNN      & KNN     & KNN       & KNN     & SVM    & KNN        & KNN     & SVM    \\
SVM       & SVM    & KNN   & SVM        & CART     & KNN     & SVM        & SVM      & SVM     & SVM       & SVM     & KNN    & SVM        & SVM     & KNN    \\
CART      & CART   & CART  & CART       & SVM      & CART    & CART       & CART     & CART    & CART      & CART    & CART   & CART       & CART    & CART 
 \\ \hline
\end{tabular}
}
 \caption{\tuned scenario.}\label{tab:tuning-ranking}
\end{subtable}
\end{table*}

\begin{table*}[tb]
\caption{RQ1-3: MAE results obtained by each of the {\sc ml} libraries for each of the five datasets for the (a) \oob scenario and the (b) \tuned scenario.} 
\label{tab:rawmae}
\begin{subtable}{1\textwidth}
\centering
\resizebox{0.95\linewidth}{!}{
\begin{tabular}{l|rrr|rrr|rrr|rrr}
\hline
                                   & \multicolumn{3}{c|}{\textbf{CART}}                                                                             & \multicolumn{3}{c|}{\textbf{KNN}}                                                                                         & \multicolumn{3}{c|}{\textbf{LR}}                                                                                          & \multicolumn{3}{c}{\textbf{SVM}}                                                                                          \\
\multirow{-2}{*}{\textbf{Dataset}} & \multicolumn{1}{c}{\textbf{SkLearn}} & \multicolumn{1}{c}{\textbf{Caret}} & \multicolumn{1}{c|}{\textbf{Weka}} & \multicolumn{1}{c}{\textbf{SkLearn}} & \multicolumn{1}{c}{\textbf{Caret}} & \multicolumn{1}{c|}{\textbf{Weka}}            & \multicolumn{1}{c}{\textbf{SkLearn}} & \multicolumn{1}{c}{\textbf{Caret}} & \multicolumn{1}{c|}{\textbf{Weka}}            & \multicolumn{1}{c}{\textbf{SkLearn}} & \multicolumn{1}{c}{\textbf{Caret}} & \multicolumn{1}{c}{\textbf{Weka}}             \\ \hline
\textbf{China}                     & 3,606.42                                & 2,963.80                              & 3,129.69                              & 2,798.93                             & 2,699.45                           & 3,532.56                                      & 2,647.40                             & 2,647.56                           & 3,067.63                                      & 3,108.56                             & 2,749.33                           & 2,471.66                                      \\
\textbf{Desharnais}                & 2,906.33                                & 2,510.66                              & 3,078.18                              & 2,199.65                             & 2,310.67                           &  3,171.84 & 2,277.26                             & 2,277.23                           &  2,051.05 & 2,754.36                             & 2,314.77                           & 2,177.32 \\
\textbf{Kitchenham}                & 2,711.59                                & 2,665.82                              & 2,484.42                              & 2,069.70                             & 2,033.88                           & 2,194.41                                      & 1,645.08                             & 1,644.95                           & 1,822.51                                      & 2,279.89                             & 2,146.26                           & 1,596.76                                      \\
\textbf{Maxwell}                   & 7,564.52                                & 4,485.85                              & 6,469.05                              & 3,850.34                             & 4,043.83                           & 6,285.81                                      & 4,448.94                             & 4,449.08                           & 3,999.02                                      & 5,741.87                             & 4,150.91                           & 4,423.29                                      \\
\textbf{Miyazaki}                  & 12,058.73                               & 13,710.94                             & 11,480.30                             & 8,644.25                             & 8,718.85                           & 9,162.33                                      & 11,700.77                            & 11,700.95                          & 12,930.09                                     & 10,346.04                            & 9,632.12                           & 7,158.33                                      \\ \hline
\end{tabular}}
\caption{\oob scenario}\label{tab:out-of-the-box-mae}
\end{subtable}

\bigskip

\begin{subtable}{0.85\textwidth}
\centering
\resizebox{0.95\linewidth}{!}{
\begin{tabular}{l|rrr|rrr|rrr}
\hline
\multirow{2}{*}{\textbf{Dataset}} & \multicolumn{3}{c|}{\textbf{CART}}                                                                             & \multicolumn{3}{c|}{\textbf{KNN}}                                                                              &  \multicolumn{3}{c}{\textbf{SVM}}                                                                              \\
                                  & \multicolumn{1}{c}{\textbf{SkLearn}} & \multicolumn{1}{c}{\textbf{Caret}} & \multicolumn{1}{c|}{\textbf{Weka}} & \multicolumn{1}{c}{\textbf{SkLearn}} & \multicolumn{1}{c}{\textbf{Caret}} & \multicolumn{1}{c|}{\textbf{Weka}} &  \multicolumn{1}{c}{\textbf{SkLearn}} & \multicolumn{1}{c}{\textbf{Caret}} & \multicolumn{1}{c}{\textbf{Weka}} \\ \hline
\textbf{China}                    & 3,681.04                             & 2,872.11                           & 3,120.91                           & 2,587.57                             & 2,597.50                           & 2,581.95                             & 3,117.41                             & 2,696.23                           & 2,519.55                          \\
\textbf{Desharnais}               & 3,535.04                             & 2,422.54                           & 3,078.18                           & 2,249.66                             & 2,220.29                           & 2,409.41                           &  2,760.77                             & 2,424.63                           & 2,342.04                          \\
\textbf{Kitchenham}               & 2,478.44                             & 2,663.92                           & 2,725.83                           & 1,985.66                             & 1,999.46                           & 1,960.95       & 2,282.58                             & 2,032.15                           & 2,045.81                          \\
\textbf{Maxwell}                  & 7,232.53                             & 4,717.38                           & 6,580.89                           & 4,285.81                             & 3,864.54                           & 4,953.90   & 5,743.15                             & 4,699.31                           & 4,005.16                          \\
\textbf{Miyazaki}                 & 10,497.33                            & 13,550.06                          & 12,368.94                          & 9,840.04                             & 8,736.13                           & 9,439.13                           &  10,347.54                            & 9,382.00                           & 8,115.88                          \\ \hline
\end{tabular}
}
 \caption{\tuned scenario}\label{tab:tuning-mae}
\end{subtable}
\end{table*}

\begin{figure*}[]
 \includegraphics[width=0.85\linewidth]{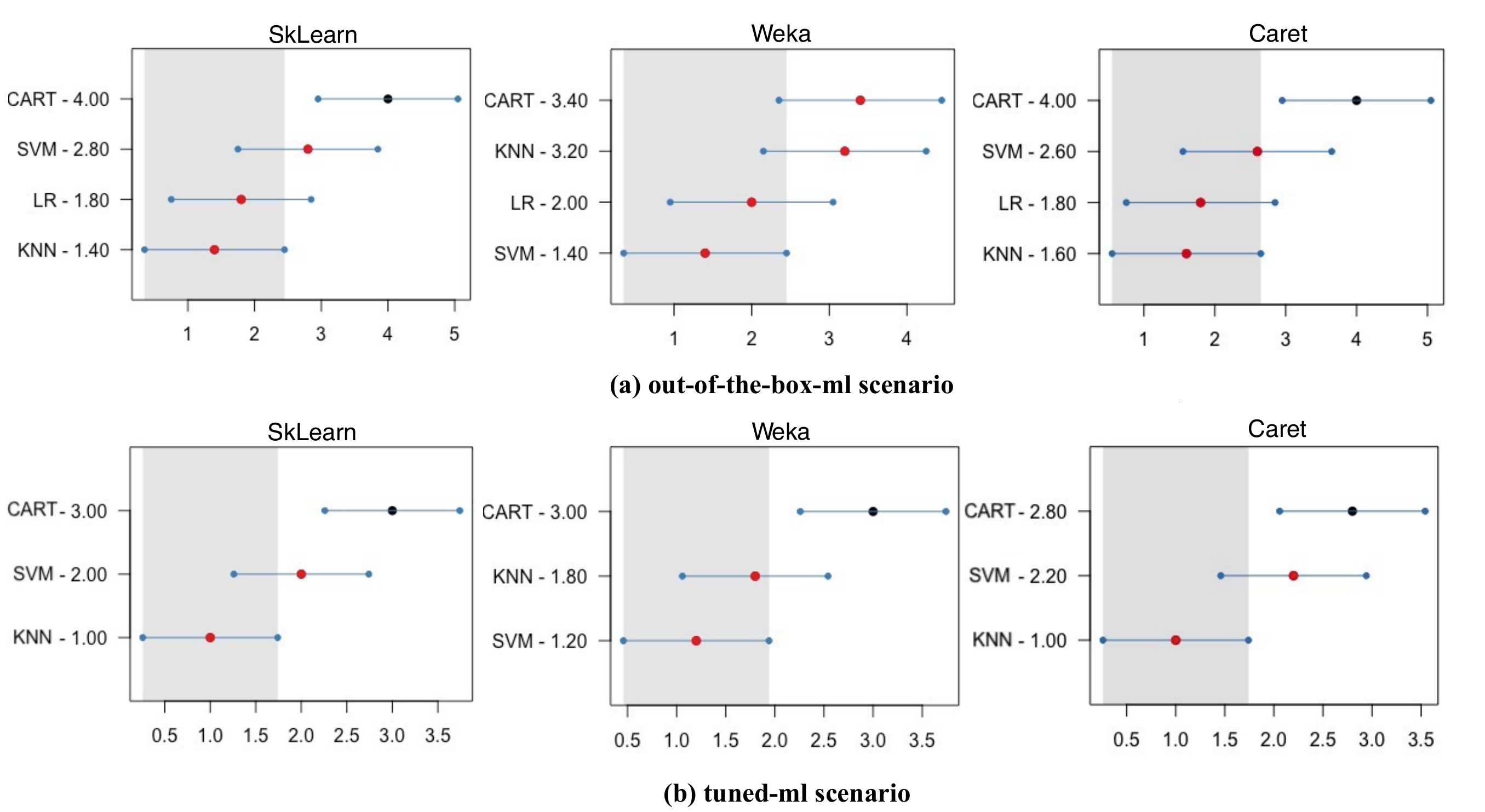}
    \caption{RQ3: Ranking of the {\sc ml} techniques based on the results of the Nemenyi Test for the (a) \oob scenario and the (b) \tuned scenario. The worst performing technique (with the highest MAE) is displayed at the top.
    }
  \label{fig:nemenyi}
\end{figure*}

The results for \oob scenario, presented in Tables \ref{tab:out-of-the-box-ranking} and \ref{tab:out-of-the-box-mae}, reveal that there is no case where all three techniques agree on a single ranking. The rankings obtained when using \weka are different from those obtained by \sk and \caret for each of the datasets investigated. Whereas, \sk and \caret provide a same ranking for three out of the five (60\%) datasets under study (i.e., China, Kitchenham, Miyazaki).
As for the \tuned scenario, we can observe that all three tools provide the same ranking on only one dataset (i.e., Kitchenham). Whereas for the remaining four, \caret and \sk agree on three datasets (i.e., China, Maxwell, Miyazaki), whereas all three libraries disagree on the fourth (i.e., Desharnais). While \sk and \caret seem to generally agree at least on some of the ranks, \weka tends to provide completely different rankings.

These observations are confirmed by the statistical significance test analysis: The Friedman test shows that the difference between the results of the techniques is statistically significant (p-value < 0.05) for all {\sc ml} libraries in both the \oob and \tuned  scenarios. A more fine grained analysis can be performed when observing the plots generated by the post-hoc Nemenyi Test. Figure \ref{fig:nemenyi} shows that, for the \oob scenario, \weka provides a different ranking than that obtained by \sk and \caret, with a statistical significant difference between the best and worst performing techniques (i.e., KNN and CART, respectively) in the case of \sk and \caret. While \sk and \caret rank KNN first and SVM third, \weka ranks KNN in third place and SVM in the first in the \oob scenario.
As for the \tuned scenario, we can observe that \weka also provides a different ranking than the other two {\sc ml} libraries (i.e., \sk and \caret) as it ranks SVM and KNN first and second, respectively. Whereas \sk and \caret both agree on ranking KNN first and SVM second. All three {\sc ml} libraries show a statistically significant difference between the best and worst ranked techniques (i.e., KNN and CART for \sk and \caret and SVM and CART for \weka).
While \sk and \caret agree on the ranking, the results of RQ2 had revealed that the MAE values achieved by these two {\sc ml} libraries are statistically significantly different. This shows that even when they output the same rankings of the {\sc ml} techniques, the results of the same techniques vary significantly depending on the {\sc ml} library used, and therefore larger estimation errors can be committed when using one library over another. 

\section{API Analysis}
\label{sec:APIanalysis}

The authors of this paper independently investigated the API documentation and source code (function body) of \caret\footnote{https://topepo.github.io/caret/}, \sk\footnote{https://scikit-learn.org/stable/modules/classes.html} and \weka\footnote{https://waikato.github.io/weka-wiki/documentation/} for each of the techniques listed in Table \ref{tab:default-param} in order to investigate similarities and differences in (1) the algorithm reference for each machine learner; (2) the signature of the main method offered by these libraries to build each machine learner (i.e., the number and type of input parameters that can be set by the user through the use of their API) and the naming convention used; (3) the parameters' default values and the values that can be assigned for categorical parameters within each library; (4) the parameters available only in the source code (i.e., \textit{hidden} parameters). We summarise our findings in Table \ref{tab:APIanalysis} and discuss them below. 

\begin{table}[]
 \caption{Summary of the API Analysis in terms of number of total parameters per {\sc ml} library and machine learner, and the number of parameters matching a same functionality, a same name, and a same default value across all libraries and each pair of libraries.}
\label{tab:APIanalysis}
\resizebox{0.7\columnwidth}{!}{
\begin{tabular}{lrrrr} 
\hline
                          & \multicolumn{1}{c}{\textit{CART}} & \multicolumn{1}{c}{\textit{KNN}} & \multicolumn{1}{c}{\textit{LR}} & \multicolumn{1}{c}{\textit{SVM}} \\ \hline
                           \multicolumn{5}{c}{Total Number of parameters}                                                                                            \\ \hline
\multicolumn{1}{l|}{\textit{Caret}}           & 9                                 & 1                                & 1                               & 3                                \\
\multicolumn{1}{l|}{\textit{SkLearn}}         & 14                                & 8                                & 4                               & 11                               \\
\multicolumn{1}{l|}{\textit{Weka}}          & 6                                 & 5                                & 0                               & 10                               \\ \hline
                         \multicolumn{5}{c}{Parameters matching functionality across}                                                                    \\ \hline
\multicolumn{1}{l|}{\textit{All libraries}}    & 2                                 & 1                                & 0                               & 2                                \\
\multicolumn{1}{l|}{\textit{SkLearn \& Weka}} & 3                                 & 3                                & 0                               & 6                                \\
\multicolumn{1}{l|}{\textit{SkLearn \& Caret}} & 4                                 & 1                                & 1                               & 3                                \\
\multicolumn{1}{l|}{\textit{Caret \& Weka}}   & 2                                 & 1                                & 0                               & 2                                \\ \hline
                          \multicolumn{5}{c}{Parameters matching name across}                                                                              \\ \hline
\multicolumn{1}{l|}{\textit{All libraries}}    & 0                                 & 0                                & 0                               & 1                                \\
\multicolumn{1}{l|}{\textit{SkLearn \& Weka}}  & 0                                 & 0                                & 0                               & 1                                \\
\multicolumn{1}{l|}{\textit{SkLearn \& Caret}} & 1                                 & 0                                & 1                               & 2                                \\
\multicolumn{1}{l|}{\textit{Caret \& Weka}}    & 0                                 & 1                                & 0                               & 1                                \\ \hline
                          \multicolumn{5}{c}{Parameters matching default values across}                                                                   \\ \hline
\multicolumn{1}{l|}{\textit{All libraries}}  & 0                                 & 0                                & 0                               & 0                                \\
\multicolumn{1}{l|}{\textit{SkLearn \& Weka}}  & 0                                 & 1                                & 0                               & 1                                \\
\multicolumn{1}{l|}{\textit{SkLearn \& Caret}} & 1                                 & 0                                & 1                               & 1                                \\
\multicolumn{1}{l|}{\textit{Caret \& Weka}}    & 0                                 & 0                                & 0                               & 0                              \\ \hline
\end{tabular}}
\end{table}

\noindent \textbf{Algorithm Reference}. We observed that the documentation of these libraries does not always clearly state a reference algorithm. Specifically, no reference was provided in seven out of the 12 cases we looked into (with \caret never providing any reference for any of the cases). We also found that, among the cases where references are given, \sk usually provides a general list of references for a given algorithm. For example, \sk provides a  list of four references for CART (including Wikipedia), without indicating the one actually being used for the implementation. Whereas among the cases where \weka provides a reference, it is a single specific one. 

\noindent \textbf{Method Signature}. 
We found that none of the libraries offer a same method signature to build each of the machine learner investigated herein (i.e., CART, KNN, LR and SVR).
All three libraries provide a different \textit{number of parameters} which can be directly manipulated by the user thorough the API as shown in Table \ref{tab:APIanalysis}. \sk provides the user with the highest number of parameters compared to \caret and \weka across all techniques. For example, in \sk, a user can build CART by specifying 14 input parameters, while \caret and \weka only allow the user to specify nine and six parameters, respectively. Similarly, a user can build KNN by specifying eight input parameters in \sk, while \caret and \weka only provide the user with the ability to specify one and five parameters, respectively. 

Among the parameters which are provided in an API but not in another, some do not directly have an effect on the accuracy of the {\sc ml} techniques, but they play a role in the execution process (e.g., $n\_jobs$ in KNN controls the number of parallel jobs to run for neighbors search); whereas others control the machine learner hyper-parameters, therefore their use/setting can directly impact the accuracy of the models.
For example, while \caret  only allows the user to set the number of neighbours to be explored for KNN, \weka permits setting the number of neighbours and the weight function, and \sk provides the user with the freedom to choose the number of nearest neighbours, the weight function of the neighbours, the algorithm used to compute the nearest neighbours, and the distance function. 

On average, across all four techniques, \sk provides the users with the ability to manipulate almost two times the number of parameters provided by \caret and three times those made available by \weka. Overall, we can state that \sk gives the user more control over the parameters and, therefore, the performance of these techniques. 

We also observe that the number of \textit{common parameters that perform the same functionality} across the three libraries is very low. 

The highest number of common parameters is only two (i.e., both CART and SVM have two common parameters, while KNN has only one and LR does not have any in common across the three libraries). If we consider each pair of libraries, we observe that \sk and \weka have more parameters in common with respect to \caret, yet the numbers remain relatively low with the best case being six parameters in common between \sk and \weka out of the 11 available ones for SVM. 

Specifically, in two cases out of 12 there is no common parameters between each pair of libraries. There is only one parameter in common in three cases, two common parameters in two cases, three common parameters in three cases. One case has four parameters in common and the remaining one has six common parameters as described above.

Even in the case where a method signature has a given parameter in common among these libraries, the parameter name, its default value, or possible categorical values, might differ across the libraries as further explained below.

\noindent \textit{Parameter Name:}  
The libraries refer to a given parameter with different names, except for one case where all three libraries use the same name. This makes it non-trivial for a user to match concepts across different libraries. For example, the parameter used to specify the distance function for the KNN model is called {\tt metric} in \sk, while in \weka it is named {\tt -A} and in \caret it is not provided. Investigating this matter between each pair of libraries, we found that in only one case, out of a total of 12, two parameters have the same name, five cases had a single parameter with a common name, while all remaining cases (i.e., six cases) did not share any parameters using a same naming convention. 

\noindent \textit{Parameter Value:} We also inspected both the documentation and source code in order to extract the number of parameters which are set to the same default values in \sk, \weka and \caret. This investigation revealed that no parameter has the same default value across all three libraries. We also compared the parameters' default values for each pair of libraries. We found that there were five cases where only one parameter had the same default value, while all the remaining cases (i.e., seven cases) did not have any parameters with the same default value. For example, \sk and \weka set a different default value in KNN for the number of nearest neighbours to be explored by the algorithm. \sk gives its {\tt n\_neighbors} parameter a value of five, while \weka sets its {\tt -K} parameter to one. Lastly, we observe that even if a given parameter is common among these libraries, they might provide the users with different value options to set categorical ones. For example, when building KNN, \weka allows the user to weigh neighbours by the inverse of their distance or by calculating $1 -$ their distance. On the other hand, \sk allows the user to choose the value of this parameter among three different options: using uniform weights, using the inverse of their distance, or by creating a user-defined function which allows the user to assign specific weights.

\noindent \textbf{Hidden Parameters}. A closer look at the implementation of these libraries revealed that some parameters that are made available to the user through API methods by one library are instead buried (i.e., \textit{hidden}) in the source code of another library. By inspecting the source code, we found that each of \weka and \caret does not directly expose to the user a number of parameters (i.e., they have hidden parameters) that could instead be configured through the API of the other two libraries. Specifically, \caret has one hidden parameter in CART and four hidden ones in SVM, while \weka has one in CART. \sk is the only library which does not have any hidden parameter that could have been matched with those made available by \weka's and \caret's APIs. 
We also discovered that some of the hidden parameters in a given library are set to values which are different from the default values set in the API of another library, and such values cannot be changed unless one modifies the source code. For example, \sk and \weka provide a parameter which defines the tolerance for the stopping criterion when building the SVR model, and both libraries setting the default value to 0.001. However, this parameter is not provided by \caret's API, instead it can only be found by investigating the code, with its value being set to 0.01 in the called function's body. 

\section{Discussion}
In this section we discuss the main findings of our study and takeaways for researchers and practitioners that use and develop {\sc ml} libraries. 

The results of our \textit{empirical study} revealed that there is a large disagreement on the effort predictions achieved by a given machine learner using these three libraries (i.e., 95\% of the cases on average across a total of 105 cases studied). Moreover, our analysis of 117 previous SEE work revealed that more than half do not state the {\sc ml} library used (51\%), and among those that do it, the majority only provides partial information (72\%), thus, making it difficult to reproduce and replicate previous work.
These findings highlight that the choice of a library is just as critical as the choice of a study's prediction techniques, benchmarks, validation procedures and evaluation measures \cite{moussa2022use}, and as such researchers should always indicate the library used in their studies.
These factors can all have an influence on a study's outcome and can contribute to the conclusion instability observed in the SEE literature. While, previous studies have highlighted the importance of the last four aspects \cite{Myrtveit05,Menzies2012,Song13,Keung2013,Sarro18,SarroTSE20}, our study is the first to investigate the impact of the {\sc ml} library.

Moreover, our \textit{analysis of the API} documentation and code suggests that software engineers building open-source {\sc ml} libraries should follow a more uniform approach by providing 
a reference to the conceptual technique implemented by a given API; exposing the right number and type of parameters needed to build a given machine learner, otherwise explain any differences; testing the implementation of a given conceptual technique which uses other existing implementations of the same technique as an oracle \cite{CynthiaOracleTestingML,zhang2020machine,BarrOracleTestingSurvey}.

To summarise, our study offers the following main \textit{take-aways} to researchers and practitioners that use and develop {\sc ml} libraries:

\begin{itemize}
\item The study raises awareness on the fact that different {\sc ml} libraries incur different results for software effort estimation. Our results show that no library, among the three studied, is best for all datasets investigated and, thus, call for actions to allow users to consider and understand the accuracy requirements when selecting an {\sc ml} library.

\item The empirical analyses we carried out can help future studies in this domain, by offering a comprehensive and sound methodology to compare {\sc ml} library performance.

\item The deficiencies we found in the current documentation of {\sc ml} libraries reveal the need for a better documentation of these libraries, and we provide some initial suggestions on how this could be tackled in future work.

\end{itemize}

\section{Threats to Validity}
The validity of our study can be affected by internal, construct, conclusion, and external threats.

\textit{Internal validity}: Our experimental setting has been mainly dictated by the need to eliminate any stochastic source yet we strove to consider datasets, regression techniques and hyper-parameter tuning widely used in previous work. 

While we acknowledge, that other experimental settings could be used to improve the overall predicting performance (e.g., using Search-based approaches for hyper-parameter tuning \cite{Corazza2010, CorazzaMFGSM13}), we explicitly avoid design choices that could introduce any stochasticity in our results, in order to soundly analyse the variances in performance of the machine learners solely due to the use of the different {\sc ml} libraries investigated in our work, rather than other possible factors.

To mitigate the threat of missing relevant information in our literature review as well as API manual analysis, two authors examined all artefacts (i.e., papers, documentation and source code) independently, in order to ensure reliability and reduce researcher bias. The results were compared at the end of the process, and any ambiguities/inconsistencies was resolved by a joint analysis and discussion.

\textit{Construct and Conclusion Validity}: 
We follow most recent best practice to evaluate and compare prediction systems \cite{ShepperdM12}. We use the MAE as a measure to evaluate and compare the predictions. The MAE is unbiased towards both over- and under-estimation and its use has been recommended \cite{ShepperdM12,Langdon2016,Sarro2018TOSEM} as opposed to other popular measures like MMRE and Pred(25) \cite{Conte:1986}, which have been criticised for being biased towards underestimations and for behaving very differently when comparing prediction models \cite{FossSKM03,Kitchenham:2001,Korte:2008,Port:2008,Shepperd2000,StensrudFKM03}. We also carefully calculated the performance measures and applied statistical tests by verifying all the required assumptions. 
We experimented with real-world datasets widely used to empirically evaluate SEE models, and to ensure a realistic scenario, we did not use any independent variables that is not known at prediction time and therefore cannot be used for prediction purposes as recommended in previous work \cite{Sarro2018TOSEM}.

\textit{External validity}: Threats related to the generalizability of our findings may arise due to the open-source libraries, techniques and datasets we investigated. We have mitigated these threats by using those that are as representative as possible of the SEE literature.

\section{Final Remarks and Future Work}
\label{sec:conclusion}
We have investigated and compared the use of three popular open-source {\sc ml} libraries (\caret, \sk and \weka) to build SEE prediction models with well-know machine learners and scenarios most commonly used in the literature.

Based on our empirical results and API analyses, we suggest and shed light on the fact that {\sc ml} library users should consider the choice of the {\sc ml} library as part of their empirical design, similar to the choice of evaluation measures or validation processes. We encourage users to justify their use of libraries which would motivate them to seek a deeper understanding of the {\sc ml} libraries and algorithms being used. On the other hand, for this to be more achievable, the documentation of such libraries should be improved to include more information about the algorithms implemented and their references. The developers of these libraries should provide a level of clarity in the documentation that would be comprehensible by all users regardless of their level of expertise.

The deficiencies we highlighted in the current API documentation, provides initial evidence of the need for further analysis of the existing APIs to derive a set of standard requirements for {\sc ml} API documentation and API construction itself, as done, for example, in previous work for the documentation of Computer Vision Software \cite{CummaudoTSE2020}.
Also, future studies can investigate the use of automated refactoring techniques to automatically recommend to developers suitable variable renaming in order to increase the consistency across different libraries. We also envisage that the use of automated deep-parameter tuning \cite{wu2015deep, bruce2016deep, bokhari2017deep} can aid to automatically improve the performance of prediction models built using those {\sc ml} libraries that do not expose as many parameters in their APIs. 
Last but not least, future work can extend our work to assess whether discordant results arise when using proprietary {\sc ml} libraries, as well as the impact of using different (open-source and proprietary) libraries for other {\sc ml}-based software engineering prediction tasks.

\section*{Acknowledgments}
Rebecca Moussa and Federica Sarro are supported by the ERC Advanced fellowship grant EPIC (741278) and the Department of Computer Science of University College London.

\balance

\bibliographystyle{ACM-Reference-Format}
\bibliography{references,tosem18}

\end{document}